*Article*

# POVMs and the Two Theorems of Naimark and Sz.-Nagy

**James D. Malley * and Anthony R. Fletcher**

Center for Information Technology, NIH, Bethesda, MD 20892, USA; E-Mail: arif@mail.nih.gov

* Author to whom correspondence should be addressed; E-Mail: jmalley@mail.nih.gov; Tel.: +1-301-496-9934.

**Abstract:** In 1940 Naimark showed that if a set of quantum observables are positive semi-definite and sum to the identity then, on a larger space, they have a joint resolution as commuting projectors. In 1955 Sz.-Nagy showed that any set of observables could be so resolved, with the resolution respecting all linear sums. Crucially, both resolutions return the correct Born probabilities for the original observables. Here, an alternative proof of the Sz.-Nagy result is given using elementary inner product spaces. A version of the resolution is then shown to respect all *products* of observables on the base space. Practical and theoretical consequences are indicated. For example, quantum statistical inference problems that involve any algebraic functionals can now be studied using classical statistical methods over commuting observables. The estimation of quantum states is a problem of this type. Further, as theoretical objects, classical and quantum systems are now distinguished by only more or less degrees of freedom.

**Keywords:** Naimark dilation; Sz.-Nagy resolution; quantum statistical inference

## 1. Introduction

For a finite dimensional quantum system in an arbitrary state consider a set of positive semi-definite observables that sum to the identity. These define a POVM, or, positive operator valued measure. Their central importance in quantum theory and practice flows from a beautiful result of Naimark, derived in 1940; see [1,2]. It shows that any POVM can be realized as commuting projectors on a larger space, and such that the projectors return the correct Born probabilities for any state of the system. Also, on this larger space the realizations of the original observables allow measurements over them without disturbing the state, and the realizations do not depend on the state. For these reasons POVMs are often assigned the coveted status of most general type of quantum measurement possible, and are often a starting point for foundational discussions in quantum information theory. Background POVM details are given in [3–17] and proofs of Naimark's result appear in [3–8].

Following Naimark's insight, Sz.-Nagy showed in 1955 how *all* observables on a finite system could be simultaneously realized as simple linear functions of commuting projectors on a single larger space; see the Appendix in [3]. It is paradoxical that this equally wonderful result of Sz.-Nagy is rarely discussed. Summarizing, from Sz.-Nagy resolutions are surrogate classical random variables with respect to all linear functions of observables. And as now shown below, the resolutions are also surrogate classical random variables with respect to all algebraic functions of observables.

In this project the result of Sz.-Nagy is re-derived and discussed using an elementary inner product space construction. It is motivated by a proof of Naimark's result as was sketched in [8], but where [8] itself makes no mention of Sz.-Nagy's result.

There are three other goals in this project.

*First*, along theoretical lines, it is shown that a version of the resolution respects *products* of arbitrary—possibly noncommuting—observables on the base space. Under this scheme all the resolutions of the observables commute, exactly as in the Naimark and Sz.-Nagy results. However, this product property is not part of either result. Therefore, as the resolution retains the algebraic structure of the observables on the base space, acquires commutativity for them, and has the correct Born probabilities, the notion of a POVM as representing the most general quantum measurement might be open to further discussion.

*Second*, along practical lines, as with the Naimark and Sz.-Nagy results, the resolution captures the marginal and conditional probabilities on the base space. This, coupled with the product property, suggests applications of classical statistical methods to quantum statistical inference problems. For example, consider a search for solutions of a functional parametric equation written over sums and products as defined by a quantum statistical inference problem on the base space. Utilizing the product property these functionals can be studied using entirely classical statistical methods over commuting observables in the resolution. Specific statistical inference examples of this process are discussed below.

Finally, *Third*, the details presented here are much more than necessary for the essential arguments. This project therefore is an effort in learning and teaching, an attempt to parse the machinery of the methods, at least for the sake of the authors. It is hoped that doing so makes these two distinctive results—these happy inventions—more transparent and accessible.

## 2. Naimark's Theorem and the Sz.-Nagy Extension

Recall that a POVM is a set of positive semi-definite observables that sum to the identity. The original result of Naimark is the scheme by which any POVM, or, generalized resolution of the identity, is realized as a set of commuting projectors on a larger space; see [1–17]. Shortly after the derivation by Naimark, an extension was presented by Sz.-Nagy [3] such that:

(i) The original observables need not be positive semi-definite or sum to the identity;
(ii) The resolution of any observable on the base system is given as a linear sum over orthogonal projectors on the larger space;
(iii) The derivation is state independent;
(iv) The resolution returns the correct Born probabilities with respect to the state of the base system.

As necessary background for the extension of Sz.-Nagy, the original Naimark result is this: *Naimark's Theorem* (1940 [1]; 1943 [2]; and see the Appendix in [3]). Suppose given a quantum system in a specific state, and a finite set of observables that are positive semi-definite and sum to the identity. Then the system can be embedded in a larger one such that the observables are simultaneously realized as commuting projectors that return the original Born probabilities with respect to the state.

A matrix-based proof of the result is outlined in [5]; an alternative matrix derivation is given in [6]; a spectral measure proof is presented in [7]; and an inner product space construction sketched in [8]. The discussion in [5] is self-contained, offering a proof of Naimark and a completely worked example for a two-dimensional base space. The proof in [6] offers an approach that enables other matrix examples of the Naimark scheme; see especially [6] (pp. 80–83).

Using the proof of Naimark given in [8], the result of Sz.-Nagy is re-derived. It leads to a proof that all *products* of observables are respected on a suitable subspace of the Naimark resolution.

The re-derivation is itself a slight extension of the Sz.-Nagy result, and is given by:

*Theorem.* Suppose given any two sets of observables $\{A_j : 1 \leq j \leq m\}$ and $\{B_i : 1 \leq i \leq k\}$ acting on a finite dimensional Hilbert space $H$. Then:

(1) There exists an embedding of $H$ in a larger Hilbert space, $\tilde{H}$, and extensions of $\{A_j\}$ and $\{B_i\}$ to operators $\{\tilde{A}_j\}$ and $\{\tilde{B}_i\}$ acting on $\tilde{H}$, such that all the $\tilde{B}_i$ commute pairwise;

(2) The embedding is an isometry and is trace-preserving in this sense:

$$tr[A_j B_i] = tr[\tilde{A}_j \tilde{B}_i], \quad 1 \leq j \leq m, 1 \leq i \leq k \tag{1}$$

*Proof.* If $\Sigma B_i \neq 1$ then introduce $B_{k+1} = I - \Sigma B_i$. If extensions $\{\tilde{B}_i : 1 \leq i \leq k+1\}$ are found such that the $\tilde{B}_i$ commute pairwise, then the same is true of the subset $\{\tilde{B}_i : 1 \leq i \leq k\}$. Moreover, given some extensions $\tilde{A}_j$, if $tr[\tilde{A}_j \tilde{B}_i] = tr[A_j B_i]$, for all $1 \leq j \leq m$, $1 \leq i \leq k+1$, then the same is true for all the $\tilde{A}_j$ and the subset $\{\tilde{B}_i : 1 \leq i \leq k\}$. Hence without loss of generality re-number the $\{B_i : 1 \leq i \leq k+1\}$ as $\{B_i : 1 \leq i \leq k\}$, and assume that $\Sigma B_i = I$.

Next, let $a$ be any real number such that $a > \max|\lambda|$, for $\lambda$ equal to any eigenvalue of any $B_i$. It follows that

$$\tilde{B}_i = (B_i + aI)/(1+ka) \quad 1 \leq i \leq k \tag{2}$$

is positive definite and $\Sigma \tilde{B}_i = I$. Again, without loss of generality, re-label every $\tilde{B}_i$ as $B_i$.

Introduce the tensor product space $\bar{H} \equiv H \otimes H_E$ with $H_E$ a Hilbert space of dimension $k$. Let any positive definite inner product on $H$ be given by $(\varphi, \zeta)$, and let $\{\omega_i\}$ be any basis for $H_E$. For $\varphi$ in $H$ define a mapping $\varphi \to \bar{\varphi}$ into $\bar{H}$ by $\bar{\varphi} = \Sigma[\varphi \otimes \omega_i] = \varphi \otimes \bar{\omega}$, where $\bar{\omega} = \Sigma \omega_i$.

Since an arbitrary element in $\bar{H}$ can be written as $\psi = \Sigma[\varphi_i \otimes \omega_i]$, for any selected set of elements in $H$, introduce the inner product on $\bar{H}$ given by

$$(\psi,\psi)_{\tilde{H}} \equiv (\tilde{\psi},\tilde{\psi}) \equiv \Sigma(\varphi_i, B_i\varphi_i) \tag{3}$$

that averages over the inner products induced by each $B_i$ on $H$. Because $\Sigma B_i = I$ this embedding of $H$ in $\bar{H}$ is such that:

$$(\tilde{\varphi},\tilde{\varphi}) = (\varphi \otimes \bar{\omega}, \varphi \otimes \bar{\omega}) = \Sigma(\varphi, B_i\varphi) = (\varphi, (\Sigma B_i)\varphi) = (\varphi,\varphi) \tag{4}$$

The embedding is therefore an isometry. Define projection operators $E_i$ acting on $\bar{H}$ by

$$E_i(\Sigma[\varphi_t \otimes \omega_t]) \equiv \varphi_i \otimes \omega_i \tag{5}$$

By inspection these are orthogonal and $\Sigma E_i = \tilde{I}$. Most importantly,

$$(\tilde{\varphi}, E_i\tilde{\varphi}) = (\varphi, E_i(\varphi \otimes \bar{\omega})) = (\tilde{\varphi}, \varphi \otimes \omega_i) = (\varphi, B_i\varphi) \tag{6}$$

so that each $E_i$ is an inner product conserving extension of each $B_i$. Next, let $A_j = (\varphi_j)(\varphi_j)^*$ be a one-dimensional observable acting on $H$, and define $\bar{A}_j$ by

$$\bar{A}_j = (\varphi_j \otimes \bar{\omega})(\varphi_j \otimes \bar{\omega})^* = (\tilde{\varphi}_j)(\tilde{\varphi}_j)^* \tag{7}$$

where the conjugate transpose is with respect to the product in $\bar{H}$ as at (4). Finally, check that for every $E_i$:

$$tr[\bar{A}_j E_i] = tr[(\tilde{\varphi}_j)(\tilde{\varphi}_j)^* E_i] = (\tilde{\varphi}_j, E_i\tilde{\varphi}_j) = (\varphi_j, B_i\varphi_j) = tr[A_j B_i] \tag{8}$$

As any operator $A_j$ can be written as a sum over projectors, by linearity the proof is complete.

The notion of *Naimark space* is next introduced, followed by the definition of *Naimark model*.

## 3. Naimark Spaces and Naimark Models

Acting on the extension space $\bar{H}$, let $N$ be defined as the family of observables that is spanned by the commutative realizations $\{\bar{B}_i\}$ of the observables $\{B_i\}$. Call space $N$ the *Naimark space* and write $N = N(\{\bar{B}_i\})$ for this set of observables. Several comments are in order.

*First*, letting $m = 1$ and $\{A_j\} = \{\rho\}$ for density $\rho$ on the base space, and assuming the set $\{B_i\}$ consists of a set of positive definite observables that sum to the identity, the *Theorem* now implies the original Naimark result. Note, importantly, the proof of the *Theorem* does not depend on the $\{A_j\}$ and thus also does not depend on the state of the base system. The flexibility in the *Theorem* for an arbitrary finite family of observables $\{A_j\}$ for $1 \le j \le k$, with $1 < k$, allows for problems that have many possible densities on the base system under consideration. An example of the need for such a family is the problem of quantum state discrimination over multiple, possible unknown states.

*Second*, suppose the $\{B_i\}$ are a spanning set of operators on $H$, but are not necessarily linearly independent, or positive semi-definite, or commuting. Since every operator on $H$ now has a linear realization over a commutative space of projectors, this implies that there always exists a single classical joint distribution function over these commuting extensions of the original observables. This joint probability distribution is exactly that given by the standard von Neumann spectral resolution result for commuting observables, and the Born marginal probabilities for the realizations agree with those in the base space.

*Third*, the two sets of observables, $\{A_j\}$ and $\{B_i\}$, in the *Theorem* do not extend in the same way. That is, arbitrary observables $B_i$ extend to linear sums over the projectors $E_i$, while observables given by states $\rho = (\varphi)(\varphi)^*$ extend to operators $(\varphi)(\varphi)^* \otimes \overline{P} = \rho \otimes \overline{P}$, for $\overline{P}$ defined as the projector on the one dimensional space spanned by $\overline{\omega}$, and by linearity otherwise. By inspection, the conclusion of the *Theorem* also obtains by using the alternative embedding that extends states $\rho = (\varphi)(\varphi)^*$ to operators of the form $(\varphi)(\varphi)^* \otimes I = \rho \otimes I$. Thus, a Naimark space will not necessarily render states as commutative in the larger space if they are given as observables in the set $\{A_j\}$, but will do so if they appear as observables in the set $\{B_i\}$.

And *Fourth*, a given Naimark space, $N = N(\{\widetilde{B_i}\})$, is not uniquely specified by the base operators $\{B_i\}$ acting on $H$, so that $N = N(\{\widetilde{B_i}\}) = N(\{\widetilde{C_{i*}}\})$, is possible with the $\{B_i\} \neq \{C_{i*}\}$.

Additional technical facts and distinctions are these:

(a) Consider the subspace in $\widetilde{H}$ given by $H \otimes \{\omega_i\}$, where $\{\omega_i\}$ is the space in $H_E$ spanned by $\omega_i$. This space is a copy in $\widetilde{H}$ of the base space $H$. Assume now that the $\omega_i$ form an orthogonal basis for $H_E$. Then the projector $P_i$ from $H_E$ onto the space $\{\omega_i\}$ can be written as $P_i = (\omega_i)(\omega_i)^*$, where the transpose here is with respect to the inner product on $H_E$. By inspection it follows that: $E_i = I \otimes P_i$;

(b) Note that the inner product on $\widetilde{H}$ is not the same as the multiplication over the separate inner products $H$ and $H_E$. That is, in general,

$$(\varphi \otimes \omega, \zeta \otimes \nu) = (\varphi \otimes \omega, \zeta \otimes \nu)_{\widetilde{H}} \neq (\varphi, \zeta)_H (\omega, \nu)_E \quad (9)$$

for $\varphi, \zeta$ in $H$ and $\omega, \nu$ in $H_E$;

(c) As further indication of the distinction just noted in (b), check that the projectors $E_i = I \otimes P_i$ as defined in the *Theorem* can also be written as:

$$E_i = \{\Sigma_i\} \otimes P_i \quad (10)$$

where

$$\Sigma_i = \Sigma(\varphi_{i(j)})(\widetilde{\varphi_{i(j)}}) \text{ and } P_i = (\omega_i)(\omega_i)^* \quad (11)$$

and where $\{\varphi_{i(j)}\}$, for $1 \leq j \leq \dim H$, is an orthogonal basis of $H$ with respect to the inner product induced by the specific observable $B_i$, so that:

$$E_i = ((\Sigma \varphi_{i(j)}) \otimes \omega_i)(\widetilde{(\Sigma \varphi_{i(j)}) \otimes \omega_i}) \quad (12)$$

(d) For any $U$ in the Naimark space, that is for $U = \Sigma \alpha_i E_i$, with complex coefficients $\alpha_i$, inspection shows that $(X \otimes I_E)U = U(X \otimes I_E)$, since $\Sigma P_i = I_E$, for $I_E$ the identity on $H_E$, and $E_i = I_H \otimes P_i$ for $I_H$ the identity on $H$.

**4. Naimark Models**

It is advantageous to formalize the embedding of a copy of the base space in the resolving space. As above, introduce a basis for $\tilde{H}$ that begins with a isomorphic copy of the base system defined as $H \otimes \{\bar{\omega}\}$, where $\{\bar{\omega}\}$ denotes the one dimensional space spanned by the vector $\bar{\omega}$ in $H_E$.

Continuing, for any operator $U$ on $\tilde{H}$, introduce the observable $U_H$ acting on $H \otimes \{\bar{\omega}\}$, where:

$$[U_H]_{ij} = [U]_{ij} \qquad 1 \le i, j \le n = \dim H \tag{13}$$

Call $U_H$ the *Naimark component* of $U$, and call the set of all such observables a *Naimark model*. The model contains, for example, the projections onto $H \otimes \{\bar{\omega}\}$ of all the observables in $N$.

Some clarifications are these:

(*a*) The term *Naimark model* is only a convenient name introduced here for operators in the Naimark space that fix the subspace $H \otimes \{\bar{\omega}\}$, and *Naimark space* is itself an introduced term. Both directly follow from the combined results of Naimark and Sz.-Nagy. Since $\dim H = \dim(H \otimes H_E)$, the space on which operators in the Naimark model act has the same dimension as the base space $H$. On the other hand the space, $H \otimes \{\bar{\omega}\}$, viewed as a subspace of $\tilde{H}$, is equipped with the inner product constructed in the *Theorem*, and this is distinct from whatever product is defined on $H$; see (*b*) in *Section* 3 above for details;

(*b*) For any observable $X$ acting on $H$, and any $U$ in the Naimark space:

$$tr[(X \otimes I)U] = tr[(X \otimes \bar{P})U] = tr[XU_H] \tag{14}$$

It follows that every observable in the Naimark model correctly returns the Born probability for the associated observable $U_H$ on the base space. This is the same probability as that given for the observable on $H$, for which observable $U$ is the Naimark resolution acting on $\tilde{H}$. In simpler terms, Naimark models are probability preserving;

(*c*) For any pair of observables on the base system, in state $D$, the quantum conditional probabilities are given by

$$\begin{aligned} \Pr[A \mid B] &= \Pr_D[A \mid B] = tr[BDBA]/tr[DB] \\ \Pr[B \mid A] &= \Pr_D[B \mid A] = tr[ADAB]/tr[DA] \end{aligned} \tag{15}$$

Consequently

$$\Pr[A \mid B] = tr[DC_{A \mid B}], \quad \Pr[B \mid A] = tr[DC_{B \mid A}] \tag{16}$$

for the two observables on the base system as defined by

$$C_{A \mid B} = BAB/tr[DB], \quad C_{B \mid A} = ABA/tr[DA] \tag{17}$$

Since any observables in the base system are expressible as simple linear sums over any spanning set, it follows that their resolutions in the Naimark model are also thus expressible, in terms of the commuting projectors in the Naimark model. In particular the pair of observables in Equation (17) have linear resolutions in the model. On the other hand, from classical probability any joint distribution on a pair of random variables is specified by the two marginal probabilities and the two conditional probabilities. Since elements of the Naimark model correctly return all marginal Born probabilities for observables on the base system, from Equations (16) and (17) it now follows that the model also correctly returns the correlation structure for any pair of observables on the base system;

(*d*) The Naimark component is defined for any operator $U$ on $\tilde{H}$ and not only for those in a Naimark space, $N$;

(*e*) For any $U$ acting on $\tilde{H}$, $(U_H\widetilde{\ })$ is always in $N$; and if $U$ is in $N$ then $(U_H\widetilde{\ }) = U$;

(*f*) For any $B$ acting on $H$: $(\tilde{B})_H = B$.

The *product property* of Naimark models is next presented.

## 5. Products and Naimark Models

Given the number, $k$, of observables in the set $\{B_i\}$, and the dimension, $m$, of the base space $H$, it is convenient to expand the size of $H$. This is most simply done by tensoring $H$ over $k$ copies of $H$. Then the dimension of the base space becomes is $t = mk$, and the dimension of $\tilde{H}$ becomes $n = mk^2$. Further, the observables $\{B_i\}$ can be trivially extended to observables of the form $\{B_i \otimes I\}$ on the expanded base space. But not so trivial is this: the extension of an observable to an operator on the expanded base space is not the same as its resolution in the Naimark space, or in the Naimark model.

One possible objective for an expansion of the base space is proposed in [8], if the dimension of the base space is not a multiple of $k$. Doing so as in [8] yields a simpler, block matrix representation of the projectors $E_i$. However, the goal of the expansion used here is different, and the utility of this particular increase in the size of $H$ is given by the following facts, later applied in *Section* 6.

Begin by letting $G$ be the projector of $\tilde{H}$ onto the base space, $H$, where that space now has adjusted dimension $t = mk$. Then, the following several conclusions obtain:

(*i*) By definition $(G)_H = I_H$, the identity on $H$. Recall that $I_H = \Sigma B_i$, for $\{B_i\}$ as in the *Theorem*. Also $(I_H\widetilde{\ }) = (\Sigma \tilde{B_i}) = \Sigma E_i = I_{\tilde{H}} = \tilde{I}$. It follows that $(G_H\widetilde{\ }) = \tilde{I} \neq G$.

(*ii*) Consider any two observables $C_1 = A_1 \otimes B_1$, $C_2 = A_2 \otimes B_2$, acting on $\tilde{H}$, such that $A_1, A_2$ act on $H \otimes \{\bar{\omega}\}$, and $B_1, B_2$ act on $H_E$. If $A_1 A_2 = A_2 A_1$, and $B_1 B_2 = B_2 B_1$ then trivially: $C_1 C_2 = C_2 C_1$.

Next, under the base space dimension adjustment just described it follows that the projector $G$ can be written as $G = Z \otimes I_k$, for a matrix $Z$, of order $mk$, having the identity matrix $I_m$ in the upper left corner, the zero matrix $(0)_t$, in the lower right corner, with $t = m(k-1)$, and zeros elsewhere. With the adjusted dimension, the projector $G$ is an operator on $\tilde{H}$, with $Z$ in the Naimark model, that is, an operator acting on $H \otimes \{\bar{\omega}\}$, a space of dimension $mk$, and with $I_k$ acting on $H_E$, a space of dimension $k$;

(*iii*) Using (*d*) in *Section* 3, and (*ii*) just given, and for any $U$ in the Naimark space $N$: $GU = UG$. Note that $G$ is not necessarily in $N$, but upon using the set $\{B_i\} = \{G, N\}$ and then applying the *Theorem*, the resolutions of $G$ and all elements of $N$ would then commute;

(*iv*) From (*iii*) just given, and for any $U, V$ in the Naimark space $N$: $(UV)_H = U_H V_H$. That is

$$(UV)_H = G(UV)G = (GUG)(GVG) = U_H V_H \qquad (18)$$

This is the *product property* of the Naimark model mentioned in the *Introduction*. It immediately extends to any linear function over finite products of observables in the model. Moreover, for *U*, *V* in *N* it is always true that $UV = VU$, and this implies: $U_H V_H = V_H U_H$;

(*v*) The extensions of the base space operators from $\{B_i\}$ to $\{B_i \otimes I_{km}\}$ work consistently with respect to the expansion of the base space. That is, a sum over operators of the form $B_i \otimes I_{km}$ has the same form, since $B = \Sigma B_i$ implies: $\Sigma(B_i \otimes I_{km}) = (\Sigma B_i) \otimes I_{km} = B \otimes I_{km}$;

(*vi*) Every observable $U_H$ in the Naimark model is by definition the Naimark component of the observable *U* in the associated Naimark space *N*. However, if an observable already fixes the base space it will not necessarily have a resolution in the Naimark space whose Naimark component is the original observable. Yet, the Naimark component will still return the correct Born probability for the original observable;

(*vii*) Given observable $U_0$ acting on the base space *H*, it follows that $\tilde{U}_0$ is in the associated Naimark space *N*, and $(\tilde{U}_0)_H$ is in the associated Naimark model;

(*viii*) Finally, every observable in the Naimark model correctly returns the Born probabilities:
$$tr[\rho(\tilde{U}_0)_H] = tr[(\rho \otimes I)\tilde{U}_0] \qquad (19)$$

## 6. Applications of Naimark Models

Consider any outcomes for any finite set of observables on the base space, $X = \{X_i\}$, and a multivariate polynomial functional equation of the form $f(X, \beta) = 0$. Using the outcomes considered as a data vector *X*, an estimated value of the parameter $\beta$ is required, such that it solves the equation to sufficient accuracy. By definition any such equation is a linear sum over products of the observables. Hence by extending the equation on the base space, the observables in the data vector are simply replaced by their resolutions in the Naimark model. And then using the product property of the extension, as above, it follows that:
$$0 = \widetilde{f(X,\beta)} = f(\tilde{X}, \beta) \qquad (20)$$

In the Naimark model the extended observables $\tilde{X}$ act as classical random variables, and have linear representations over a set of commuting projectors. Therefore, if a statistical solution with sufficient accuracy can be found in the Naimark model, then the product property, result (*iv*) in *Section* 5 above, can be applied so that exactly, or at least approximately:
$$0 = f(\tilde{X}, \beta)_H = f(\tilde{X}_H, \beta) = f(X, \beta) \qquad (21)$$

Significant to note here is that any polynomial equation over the data in the Naimark model, obtained by just adding degrees of freedom as in result (*v*) of *Section* 5 above, reduces to a solution of the equation with the same form in the base space. That is:
$$0 = f(X \otimes I, \beta) \quad \text{if and only if} \quad 0 = f(X, \beta) \otimes I \quad \text{if and only if} \quad 0 = f(X, \beta) \qquad (22)$$

In still other words, the Naimark model respects all *algebraic* constraints posed by functional equations on the base space, and the observables resolved in the Naimark model are all commuting.

As one example of this process consider using the Expectation-Maximization (EM) algorithm, a scheme from classical statistics, for study of neutron absorption tomography, as given in [12] (Section 10.3).

An alternative path here is introduction of the Naimark model, followed by application of the classical EM algorithm to the set of classical random variables that represent the commuting observables in the model. For more detail on the classical EM algorithm itself see, for example ([18] Chapter 4).

As another example of a quantum statistical problem that starts from an equation of the form $f(X,\beta) = 0$ on the base space, consider quantum state discrimination, or, state estimation. Such is presented in a Bayesian solution derived by Holevo, and Yuen *et al*. For a detailed discussion see [5,12].

Important to notice here is that a Naimark model also respects any required algebraic side conditions necessary for a quantum statistical problem that involve the original—or possibly additional—observables in the data vector. And this is exactly the case in the solution for the Bayesian state estimation problem just mentioned, where a certain semi-positivity side condition for these observables on the larger Naimark model is required; see [5,12]. Using the commutativity of the observables realized over the Naimark model, and since the observables in the Naimark space project onto the Naimark model, the resolving observables in the Naimark model must also respect the semi-positivity condition.

A more detailed resolution of this problem of Bayesian quantum state estimation using a Naimark model will be given elsewhere; see [19].

Continuing, two other classical, and widely deployed statistical estimation techniques are generalized linear models and generalized estimating equations, about which see [20]. In these schemes the functional equation $f(X,\beta) = 0$ is replaced by:

$$f(Y, X, \beta) = 0 \qquad (23)$$

where *Y* is an outcome for which a good approximation or prediction is sought, using observations *X* as mediated by a parameter $\beta$, that is to be estimated in (23). Both these methods are now, in principle, applicable to quantum estimation problems and such that a classical Naimark model solution over classical random variables is then obtained.

Another application of the *Theorem* is this. Consider given two POVMs, $\{P_i\}$ and $\{Q_j\}$, where the number of observables in each set need not be the same in number. In [11] the following problem is stated: is it possible to identify a single POVM, $\{R_i\}$, such that its elements contain all members of the original two POVMs?

If the purpose of any single POVM is to describe a set of observables having a resolution as commuting observables on a larger system, then the *Theorem* already provides that. In this case the resolution over the union of the observables in the two POVMs on a larger system jointly resolves all elements of the two POVMs as commuting observables.

On the other hand, if the task is to identify a single POVM on the base space that contains both sets of observables, then the following can be applied. As each POVM separately sums to the identity, the sum over all the observables in both POVMs must sum to twice the identity. Now divide all observables by ½. Then the full list of observables sums to the identity, and the *Theorem* applies. The result is a resolution that, apart from the factor ½, jointly returns the Born probabilities and is composed of commuting observables on the larger space. The central point here is that no loss in generality of quantum measurement is incurred in this scheme.

Finally, here is a method that uses only the original version of Naimark ([1,2]), but now applied twice. Let the first POVM be labeled, *P* and the second *Q*. Using the original Naimark result embed the original space in a larger one, such that all observables in the extensions of elements of *P* are commuting among themselves and sum to the identity on the larger space. As usual the extensions of the elements of *P* now sum to the identity on the larger space. Further, the elements of *Q*, as with all other observables apart from those in *P*, will extend in the usual way to observables that are tensor products of the elements of *Q*, by the identity on the larger space.

At this point inspection of the Naimark result shows that all elements of the extensions of *P* and *Q* must commute. That is the extensions of *P* are linear sums over projectors on the larger space and these commute with the extensions of *Q* that are simple tensors by the identity; see the *Third* note in *Section* 3.

However, at this stage, elements of the extensions of *Q* need not be commuting among themselves. Thus, for the next step, apply the classical Naimark result to the observables in the extension of *Q* on the larger space. The extensions of the original observables in *P*, on the larger space, will extend in the usual way for tensor products, to observables on the still larger space that also commute, since they did so on the larger space. Finally, note that the doubly extended forms of *P* and *Q* must jointly commute on the double extension, and the two together sum to the identity.

In effect, this method uses the *Theorem* above, where elements of the POVMs, *Q* and *P* are respectively the subsets $\{A_j\}$ and $\{B_i\}$.

## 7. A Theoretical Perspective for Naimark Spaces

The construction of the resolving Naimark space shows that it, itself, lives in the larger space where—also by construction—there are observables that are not necessarily commuting. Hence every such Naimark resolution is nested in a space that again reveals quantum and not just classical behavior. In turn, the family of all observables in the resolving space is in principle realizable in a still larger space, for which the resolved observables would all commute and act classically. Yet this entails introduction of still other noncommuting observables, thus continuing the sequence.

The scheme presented here using Naimark spaces and models suggests that for finite systems there may not be a sharply defined or even usefully declared boundary between quantum and classical phenomena. If a system is seen as classical in one Naimark representation then there are necessarily other observables in the embedding space that are noncommuting and show quantum behavior. In other words, classical and quantum phenomena are nested within each other.

On this view, quantum systems differ from classical ones simply by having different degrees of freedom. Further, the introduction of Naimark models shows something else. The Naimark model has the same dimension of the base space and this is strictly less than the dimension of the supervening Naimark space. Also, the observables on the base space, as trivially extended to those on the Naimark model are now all commuting. Finally, all algebraic functionals or constraints given over observables on the base space remain valid in the model.

In words, algebraic functionality on the base space is invariant with respect to its representation as a Naimark model, and the Naimark model observables are commuting and in this sense act classically.

## 8. Conclusions

The introduction of Naimark spaces leads to a canonical procedure by which any finite dimensional quantum system can be rendered as a classical system, such that it appears as a subspace of a larger quantum system, and such that all the original observables now have commutative versions, and possess a single classical joint distribution. Notably, under the extension it is not required that the original observables on the base space be projectors, or positive semi-definite, or sum to the identity, or be commuting. This much is all valid using the original results of Naimark and Sz.-Nagy.

As shown above the resolving Naimark model also respects algebraic constraints over any observables on the base space, and this property is not part of the original Naimark result or the Sz.-Nagy extension. Thus, the family of jointly measureable observables in a Naimark model can resolve some forms of quantum statistical estimation and detection problems, but now using entirely classical procedures over commuting observables.

Naturally any experimental implementation of the results of Naimark and Sz.-Nagy, and solutions found in a Naimark model, may not be a feasible. Hence for some problems, the resolving classical systems and the derived classical inference solutions might remain theoretical constructions. Still, the increasing use of so-called ancilla and ancillary systems suggests that the engineering task of invoking Naimark models is increasingly cost efficient.

Finally, the schemes presented here suggest that a sharp boundary between classical and quantum systems is more porous and more fluid than suspected, as the systems are nested within each other and distinguished only by counting degrees of freedom. In still other words, quantum behavior arises by restriction of classical behavior, and classical systems lurk within quantum ones, and only an engineering task separates them.

## Acknowledgments

This work was supported by the Intramural Research Program of the National Institutes of Health.

## Conflicts of Interest

The authors declare no conflicts of interest.